\documentclass[aps,prl,twocolumn,superscriptaddress,nofootinbib]{revtex4-2}
\usepackage[english]{babel}
\usepackage{bm}
\usepackage{amsmath}
\usepackage[dvips]{graphicx}
\usepackage[colorlinks=true, allcolors=magenta]{hyperref}
\usepackage{diagbox}
\usepackage{multirow}
\usepackage{threeparttable}
\usepackage{xfrac}
\DeclareMathOperator{\arccot}{arccot}

\begin{document}
\title{Dislocations in twistronic heterostructures}

\author{V.~V.~Enaldiev}
\affiliation{Abrikosov Center for Theoretical physics, MIPT, Dolgoprudnyi, Moscow Region 141701, Russia} 
\affiliation{Kotelnikov Institute of Radio-engineering and Electronics of the RAS, Mokhovaya 11-7, Moscow 125009, Russia}

\begin{abstract}
Long-period moir\'e superlattices at the twisted interface of van der Waals heterostructures relax into preferential-stacking domains separated by dislocation networks. Here, we develop a mesoscale theory for dislocations in the networks formed in twistronic bilayers with parallel (P) and antiparallel (AP) alignment of unit cells across the twisted interface. For P bilayers we find an exact analytical displacement field across partial dislocations and determine analytic dependences of energy per unit length and width on orientation and microscopic model parameters. For AP bilayers we formulate a semi-analytical approximation for displacement fields across perfect dislocations, establishing parametric dependences for their widths and energies per unit length. In addition, we find regions in parametric space of crystal thicknesses and moir\'e periods for strong and weak relaxation of moir\'e pattern in multilayered twistronic heterostructures. 
\end{abstract}
\maketitle

Study of van der Waals heterostructures, featuring moir\'e superlattice at the twisted interface between constituent layers, became a rapidly growing research area boosted by discovery of novel quantum phenomena \cite{cao2018correlated,cao2018unconventional,Yankowitz2019,Xu2019,yoo2019atomic,lu2019superconductors,sharpe2019emergent,cao2021nematicity,Wang2020,Ghiotto2021,stern2021,woods2021,yasuda2021,wang2021,Weston2022}. Moir\'e superlattices in the twistronic structures are due to periodic spatial variation of local stacking arrangement, with a period determined by misorientation angle and/or lattice mismatch between layers at the interface. Long period moir\'e superlattices, specific for small-angle interfacial twist, minimize energy of lattice transforming moir\'e pattern into arrays of preferential stacking domains separated by network of domain walls \cite{AldenPNAS,yoo2019atomic,rosenberger2020,Weston2020,McGilly2020,Halbertal2021,Shabani2021,Kazmierczak2021,Kerelsky2021,Zhang2022}. 

\begin{figure}[t]
	%\centering
	\includegraphics[width=1.0\columnwidth]{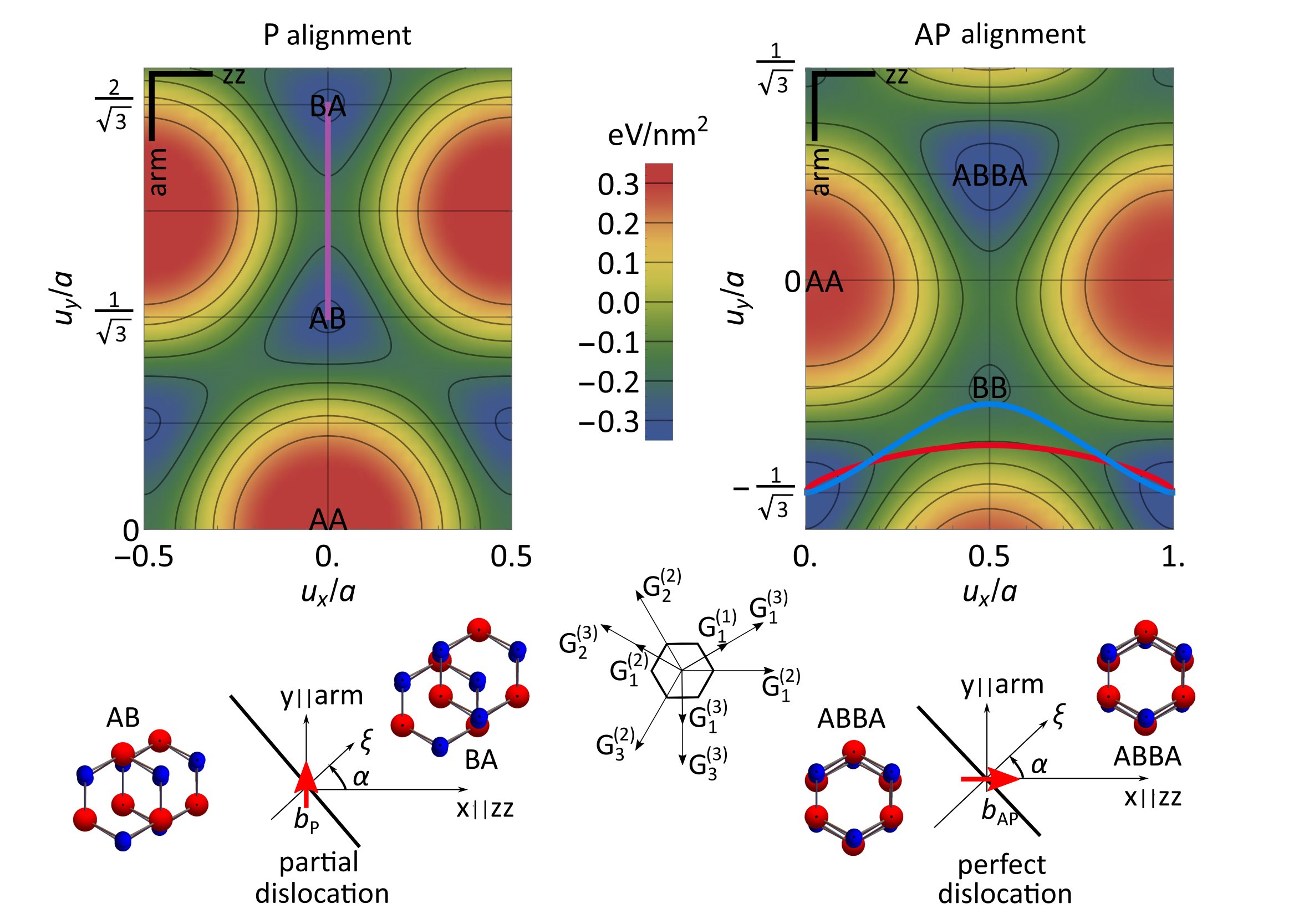}
	\caption{\label{fig:0} (Left) Map of adhesion energy densities \eqref{Eq:2} for twistronic bilayers with P alignment. Non-equivalent minima of the adhesion energy occur at $\bm{u}=(0,a/\sqrt{3})$ (AB stacking) and $\bm{u}=(0,2a/\sqrt{3})$ (BA stacking). Purple line indicates path across a partial dislocation with Burgers vector $\bm{b}_{\rm P}=(0,a/\sqrt{3})$. (Right) same for AP twistronic bilayers. Energetically favorable ABBA stacking corresponds to $\bm{u}=(0,-a/\sqrt{3})$. Red/blue line highlights path across a screw/edge perfect dislocation with $\bm{b}_{\rm AP}=(a,0)$. Middle inset sketches reciprocal vectors used in Eq. \eqref{Eq:2}. On bottom insets, 'zz' and 'arm' designate zigzag and armchair crystallographic directions in layers, respectively. }
\end{figure}

Vertical assembly of twistronic heterostructures, composed of non-inversion symmetric blocks, such as transition metal dichalcogenide (TMD) or hexagonal boron nitride (hBN) monolayers, results in different domain structures for parallel (P) and antiparallel (AP) alignment of unit cells across the twisted interface. For P alignment, relaxed moir\'e pattern consists of triangular domains characterized by the lowest energy AB (BA) stacking \cite{rosenberger2020,Weston2020,Halbertal2021,Liang2023}, that corresponds to vertical alignment of A (B) and B (A) sublattices in two layers at the interface (see Fig. \ref{fig:0}). We accept stacking nomenclature used for graphene-based heterostructures, where AB and BA stackings correspond to alignment of Bernal graphene bilayers \cite{AldenPNAS,Butz2013,yoo2019atomic,Annevelink2020}, whereas for TMD structures the stackings possess atomic structure of rhombohedral polytype \cite{CarrPRB2018,NaikPRL2018,Enaldiev_PRL,Strachan2021}.

For twistronic heterostructures with AP alignment, relaxed moir\'e pattern consists of hexagonal domains with energetically favourable ABBA-stacking \cite{Weston2020,Halbertal2021}, characterized by simultaneous vertical overlap of A and B sublattices in one layer with B and A sublattices of neighboring layer across the interface. Such ABBA stacking, shown on the inset in Fig. \ref{fig:0}, is identical to alignment of adjacent layers in bulk hexagonal TMD and BN crystals \cite{Enaldiev_PRL,Ribeiro2011,Zhou2015}.  

%%%%%%%%%%%%%%%%%%%%%%%%%%%%%%%%%%%%%%%%%%%%%%%%%%%%%%%%%%%%%%%%%
%%%%%%%%%%%%%%%%%%%%%%%%%%%%%%%%%%%%%%%%%%%%%%%%%%%%%%%%%%%%%%%%%

Domain walls separating preferential stacking domains in P (AP) twistronic heterostructures can be identified with extended crystalline defects known as partial (perfect) dislocations \cite{Enaldiev_PRL,Annevelink2020,Engelke2023}, characterized by Burgers vector $|\bm{b}_{\rm P}|=a/\sqrt{3}$ ($|\bm{b}_{\rm AP}|=a$). In case of equilateral domains (triangular for P and hexagonal for AP) forming at the interface with undistorted moir\'e pattern, the dislocations have screw type, with the Burgers vectors aligned with dislocation lines, whereas, in general case, the partial/perfect dislocations possess both screw and edge components of displacement field \cite{Butz2013,Weston2020,Engelke2023}.

Here, we develop a theory for partial/perfect dislocations forming in small-angle P/AP twistronic bilayers of graphene, hBN and MX$_2$ TMD (M=Mo, W; X=S, Se), combining elasticity theory and microscopic models for adhesion energy between constituent layers. For P bilayers we find an exact analytic distribution for displacement field across partial dislocations determining explicit dependences of their widths and energies per unit length on orientation and microscopic model parameters. Using these characteristics and calculating minimal size of a seed as a function of repolarised electric field, we consider several scenarios for ferroelectric polarisation switch in P TMD and hBN bilayers exhibiting interfacial ferroelectricity. For AP bilayers we formulate a semi-analytical approximation for displacement field across perfect dislocations providing parametric and orientation dependences of widths and energy per unit length, which are compared with the exact ones. Using an energetical criterium we find out dependence of  minimal moir\'e superlattice period {\it versus} number of layers for transformation of moir\'e pattern into domain structures in multilayered twistronic heterostructures.    

{\bf Mesoscale model for dislocations.} Using smoothness of the dislocations on atomic scale \cite{Weston2020,Edelberg2020} we formulate a continuous mesoscale theory based on the energy functional, 
\begin{equation}\label{Eq:1}
	\mathcal{E}=\int_{-\infty}^{\infty}\left[W_{\rm ad}(\bm{u}) + W_{\rm el}(u_{ij})\right]d\xi,
\end{equation}  
accounting for sum of elastic and adhesion energy densities across dislocation axis ($\xi=0$) set by an angle $\alpha$ ($-\pi/2\leq\alpha\leq\pi/2$) with respect to zigzag crystallographic direction (see Fig. \ref{fig:0}). The adhesion energy density \cite{Zhou2015,CarrPRB2018,Enaldiev_PRL}, 
\begin{align}\label{Eq:2}
	%\begin{split}
	W_{\rm ad}(\bm{u}) =% \qquad\qquad\qquad\qquad\qquad\qquad\qquad\qquad\qquad\\
	\sum_{\substack{n=1,2,3\\l=1,2,3}}\left[w^{(s)}_{n}\cos\left(\bm{G}^{(n)}_l\bm{u}\right) + w^{(a)}_{n}\sin\left(\bm{G}^{(n)}_l\bm{u}\right)\right],
	%\end{split}
\end{align}
is determined by an interlayer lateral offset, \mbox{$\bm{u}=(u_x,u_y)$}, counted from AA stacking for P and AP bilayers, and three smallest triads of reciprocal lattice vectors, $|\bm{G}^{(1)}_{1,2,3}|=4\pi/a\sqrt{3}\equiv G$, $|\bm{G}^{(2)}_{1,2,3}|=G\sqrt{3}$, $|\bm{G}^{(3)}_{1,2,3}|=2G$, with orientation shown on middle inset in Fig. \ref{fig:0}. For P bilayers this choice of origin for the interlayer offsets leads to \mbox{$w^{(a)}_{1,2,3}=0$} \cite{CarrPRB2018,Enaldiev_PRL} and, consequently, $W_{\rm ad}(\bm{u})=W_{\rm ad}(-\bm{u})$, with two inequivalent minima -- AB and BA. The elastic energy density,  
\begin{equation}\label{Eq:elen}
	W_{\rm el} = \frac{\lambda}{4}u_{ii}^2+\frac{\mu}{2}u_{ij}u_{ji},  
\end{equation}
is determined by strain tensor $u_{ij}=\tfrac12\left(\tfrac{\partial u_j}{\partial x_i}+\tfrac{\partial u_i}{\partial x_j}\right)$ and Lam\'e parameters, $\lambda$ and $\mu$, of a monolayer. Note that $u_{ij}$ characterizes relative strain between the two layers which is incorporated in Eq. \eqref{Eq:elen} by halving elastic energy of a single layer. Material-dependent energy parameters in Eq. \eqref{Eq:1} are gathered in Table \ref{tab_parameters}.

\begin{table}[t]
%	\resizebox{\columnwidth}{!}{
		\caption{Elastic and adhesion parameters (in eV/nm$^2$) for studied materials. Last column shows values of fitting $\eta$  in Eq. \eqref{Eq:exactPenergy}. \label{tab_parameters}}
		\begin{threeparttable}
			\begin{tabular}{lc|ccccccc|c}
				\hline
				\hline
				& &  $\lambda$  & $\mu$ & $w_{1}^{(s)}$ & $w_{2}^{(s)}$ & $w_{3}^{(s)}$ & $w_{1}^{(a)}$, & $w_{3}^{(a)}$, & $\eta$  \\ 
				& &  &  & $\times10^{-3}$ & $\times10^{-3}$ & $\times10^{-3}$& $\times10^{-3}$ & $\times10^{-3}$ & \\
				\hline 
				\multirow{2}{*}{MoS$_2$}& P &  \multirow{2}{*}{520} & \multirow{2}{*}{443} & 151.9 & -3.71 & -1.85 &  &  & 1.9 \\ 
				& AP & &  & 138.9 & -3.19 & -1.57 & 23.8 & -0.26 &     \\
				\hline
				\multirow{2}{*}{WS$_2$}& P &\multirow{2}{*}{328} & \multirow{2}{*}{453} & 166.4 & -4.8 & -2.4 &  &  & 2.05 \\ 
				 & AP &  &    &  151.7   & -4.09  & -2.01   &  29.55  &  -0.36 &   \\	 
				\hline
				\multirow{2}{*}{MoSe$_2$}& P&  \multirow{2}{*}{260} & \multirow{2}{*}{306} & 155.3 &  -4.25 & -2.12 &  &  & 2.017 \\ 
				& AP & &  & 138.6 & -3.5 & -1.71 & 32.51 & -0.37 &   \\	
				\hline
				\multirow{2}{*}{WSe$_2$} & P &  \multirow{2}{*}{185} & \multirow{2}{*}{303} & 132.6 &  -2.75 & -1.38 &  & & 1.76 \\ 
				& AP & &  & 112.6 & -2.12 & -1.01 & 33.63 & -0.31 &   \\
				\hline
				\multirow{2}{*}{hBN}& P & \multirow{2}{*}{627} & \multirow{2}{*}{736} & 89.5 &  -0.31 & 2.11 &  &  &  1.455 \\ 
				& AP & &  & 62.1 & -0.57 & -0.2 & 49.05 & -0.42 &  \\
				\hline
				Gr& & 617 & 1007 & 77.5 &  -0.71 & -0.18 &  &  & 1.455 \\
				\hline
				\hline
				%\multicolumn{9}{l}{$^*)$ For $\alpha^{\rm 3R}_{zz}$, $\chi$ and $\epsilon_{zz}$ we show the values obtained with}
			\end{tabular}
		\end{threeparttable}
%	}
\end{table}
%%%%%%%%%%%%%%%%%%%%%%%%%%%%%%%%%%%%%%%%%%%%%%%%%%%%%

\begin{figure}[t]
	%\centering
	\includegraphics[width=1.0\columnwidth]{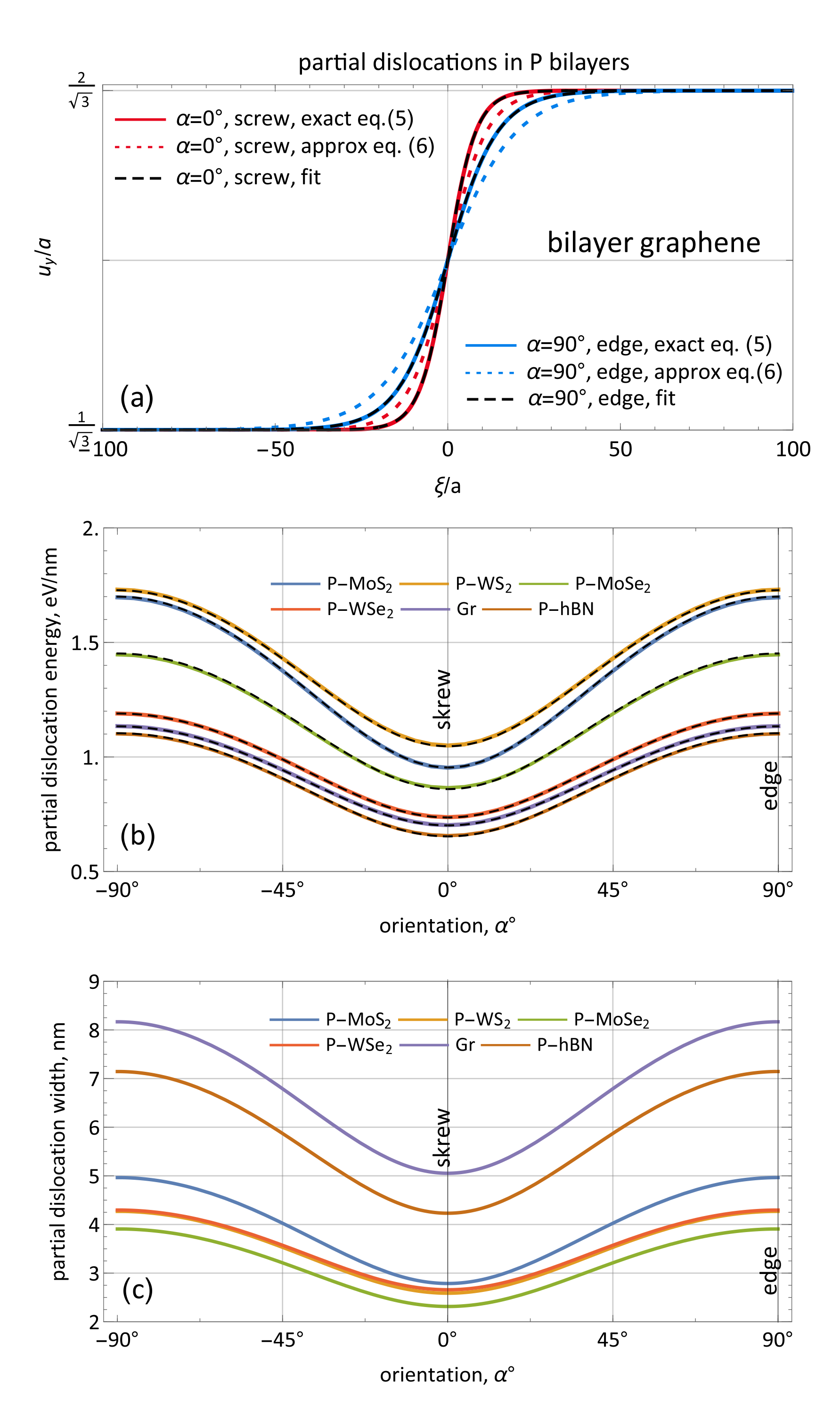}
	\caption{\label{fig:P_DW} (a) Exact, Eq. \eqref{Eq:4} (solid), approximated, Eq. \eqref{Eq:5} (dotted) and fitted (black dashed) displacement fields across screw and edge partial dislocations in bilayer graphene. Fits are made with Eq. \eqref{Eq:5} and rescaling $w^{(s)}_{1}\to \eta w^{(s)}_{1}$. (b) Orientation dependences of partial dislocation energies per unit length. Dashed lines show fits of the exact values with $\mathcal{E}_{\rm P}(\alpha)$, Eq. \eqref{Eq:exactPenergy}. (c) Orientation dependence of partial dislocation widths, $\mathcal{W}_{\rm P}(\alpha)/\sqrt{\eta}$.}
\end{figure}

{\bf First, we consider a partial dislocation}, characterized by a Burgers vector $\bm{b}_{P}=(0,a/\sqrt{3})$, in P bilayers (see inset in Fig. \ref{fig:0}). Since crossover between AB and BA domains occurs along a straight line \cite{Popov2011,Lebedeva2019,Enaldiev_PRL}, depicted by purple in Fig. \ref{fig:0}, displacement field, $\bm{u}=(0,u_y)$, across the partial dislocation is determined by the only component parallel to $\bm{b}_{\rm P}$. To determine $u_y(\xi)$ we minimize functional, Eq. \eqref{Eq:1}, which leads to the following equation,
\begin{multline}\label{Eq:3}
	\frac{\left[\mu+\left(\lambda+\mu\right)\sin^2\alpha\right]}{-4G}\frac{\partial^2u_y}{\partial\xi^2} = w_1^{(s)}\sin\left(\tfrac{3Gu_y}{4}\right)\cos\left(\tfrac{Gu_y}{4}\right)\quad\\
	 +\sin\left(\tfrac{3Gu_y}{2}\right)\left[2w_3^{(s)}\cos\left(\tfrac{Gu_y}{2}\right)+\tfrac{3}{2}w_2^{(s)}\right],
\end{multline}
%$u_{y}(-\infty)=a/\sqrt{3}$ and $u_y(+\infty)=2a/\sqrt{3}$,
with boundary conditions, $u_y(-\infty)=a/\sqrt{3}$ and $u_y(+\infty)=2a/\sqrt{3}$, accounting for crossover from AB (at $\xi=-\infty$) to BA (at $\xi=+\infty$) stacking domains (see Fig. \eqref{fig:0}). Integrating Eq. \eqref{Eq:3} with the help of a table integral \cite{Prudnikov_Marichev}, one finds an analytic solution:
\begin{multline}\label{Eq:4}
	\frac{\pi\xi}{a}\left(r_- +1\right)\sqrt{\frac{-8w_3^{(s)}\left(1+\frac23r_+\right)\left(1-2r_-\right)}{\mu+\left(\lambda+\mu\right)\sin^2\alpha}} = \\
	{\rm sign}(2u_y-a\sqrt{3})\left[(3+2r_-)\Pi(\varphi,\nu, k)-{\rm F}(\varphi, k)\right].
\end{multline}
Here, ${\rm F}(\varphi,k)$ and $\Pi(\varphi,\nu,k)$ are the incomplete elliptic integrals of the first and third kinds, with the following arguments:
\begin{align}
	\varphi =& \arcsin\sqrt{\frac{\left(1-2r_-\right)\left[\cos\left(\tfrac{Gu_y}{2}\right)+1 \right]}{4\cos\left(\tfrac{Gu_y}{2}\right)-2-4r_- }}, \nonumber\\
	\nu =& \frac{8(r_-+1)}{2r_- -1}, \nonumber \\
	k=&\sqrt{\frac{8\left(r_+-r_-\right)}{(3+2r_+)(1-2r_-)}}, \nonumber
\end{align}
and $$r_{\pm}=-\tfrac{w^{(s)}_2}{2w^{(s)}_3}\pm\tfrac{1}{2}\sqrt{\tfrac{2w^{(s)}_2}{w^{(s)}_3}-\tfrac{w^{(s)}_1}{w^{(s)}_3}+\left(\tfrac{w^{(s)}_2}{w^{(s)}_3}\right)^2}.$$ 
Displacement field, Eq. \eqref{Eq:4}, across a partial dislocation, exemplified in Fig. \ref{fig:P_DW}(a) for bilayer graphene, demonstrates growth of the dislocation width from screw ($\alpha^{\circ}=0^{\circ}$) to edge ($\alpha^{\circ}=90^{\circ}$) orientations. Similar width dependences exhibit partial dislocations in the other materials, indicating increase of lattice rigidity from shear to hydrostatic stresses, specific for screw and edge dislocations, respectively. Moreover, such a behaviour agrees with orientation dependences of energy per unit length of partial dislocations, shown in Fig. \ref{fig:P_DW}(b), where screw possess the minimal energy compared to the most energetically expensive edge ones. These energy dependences are obtained by numerically computing integral in Eq. \eqref{Eq:1} with the exact displacement field, Eq. \eqref{Eq:4}. Among studied materials, partial dislocations in P WS$_2$ have the highest energies per unit length, whereas for P hBN bilayers the dislocation formation costs the least energies. 

To determine analytic dependences of the dislocation energies on microscopic parameters and orientation, we obtain an approximated solution of Eq. \eqref{Eq:3} in the limit $w^{(s)}_{2,3}\to0$ \cite{Lebedeva2019}:
\begin{equation} \label{Eq:5}
	u_y(\xi)=a\sqrt{3}\left\{\tfrac{1}{2}+\tfrac{1}{\pi}\arctan\left[\tfrac{\tanh\left(\frac{2\xi}{\mathcal{W}_{\rm P}(\alpha)}\right)}{\sqrt{3}} \right] \right\},  
\end{equation} 
where, 
\begin{equation}\label{Eq:width}
	\mathcal{W}_{\rm P}(\alpha)=\frac{2a}{\pi}\sqrt{\frac{\mu+(\lambda+\mu)\sin^2\alpha}{2w_1^{(s)}}},
\end{equation}
is the partial dislocation width. In Fig. \ref{fig:P_DW}(a) we show the approximated displacement field, Eq. \eqref{Eq:5}, across screw (red dotted) and edge (blue dotted) partial dislocations in bilayer graphene which demonstrate a reasonable agreement with the exact distributions, Eq. \eqref{Eq:4}, in virtue of $w^{(s)}_{1}\gg |w_{2,3}^{(s)}|$ (see Table \ref{tab_parameters}). Moreover, we find that rescaling the adhesion parameter, $w^{(s)}_{1}\to \eta w^{(s)}_{1}$, in Eqs. \eqref{Eq:5}, \eqref{Eq:width} with a single orientation-independent parameter, $\eta$, one attains a perfect overlap with the exact displacement field, Eq. \eqref{Eq:4}, for any orientation, see black dashed lines in Fig. \ref{fig:P_DW}(a). The values of $\eta$ for each bilayer are obtained from fit of the exact energy dependences (Fig. \ref{fig:P_DW}(b)) with the help of 
\begin{equation}\label{Eq:exactPenergy}
	\mathcal{E}_{\rm P}(\alpha)=\tfrac{3a}{\sqrt{2}\pi}\left(1-\tfrac{\pi}{3\sqrt{3}}\right)\sqrt{\eta w_{1}^{(s)}\left[\mu+(\lambda+\mu)\sin^2\alpha\right]},
\end{equation}
which is obtained from functional in Eq. \eqref{Eq:1} for displacement field, Eq. \eqref{Eq:5}, with $w^{(s)}_{1}\to \eta w^{(s)}_{1}$. Results of the fitting are gathered in Table \ref{tab_parameters} and shown by black dashed in Fig. \ref{fig:P_DW}(b), where one finds a perfect matching between $\mathcal{E}_{\rm P}(\alpha)$, Eq. \eqref{Eq:exactPenergy}, and exact numerical data for every studied material. Therefore, using Eq. \eqref{Eq:5} we can identify $\mathcal{W}_{\rm P}(\alpha)/\sqrt{\eta}$ with the widths of exact partial dislocations, Eq. \eqref{Eq:3}, plotted in Fig. \ref{fig:P_DW}(c). Confirming the above-mentioned orientation dependences, the displayed width behaviours demonstrate that partial dislocations are the thickest for bilayer graphene, as the latter possess the highest lattice rigidity and the lowest adhesion (see Table \ref{tab_parameters}). We also note that the obtained $\mathcal{W}_{\rm P}(\alpha)/\sqrt{\eta}$-dependence for bilayer graphene is in a good quantitative agreement with that of extracted experimentally in Ref. \cite{AldenPNAS}. 

%%%%%%%%%%%%%%%%%%%%%%%%%%%%%%%%%%%%%%%%%%%%
\begin{figure}%[t]
	%\centering
	\includegraphics[width=1.0\columnwidth]{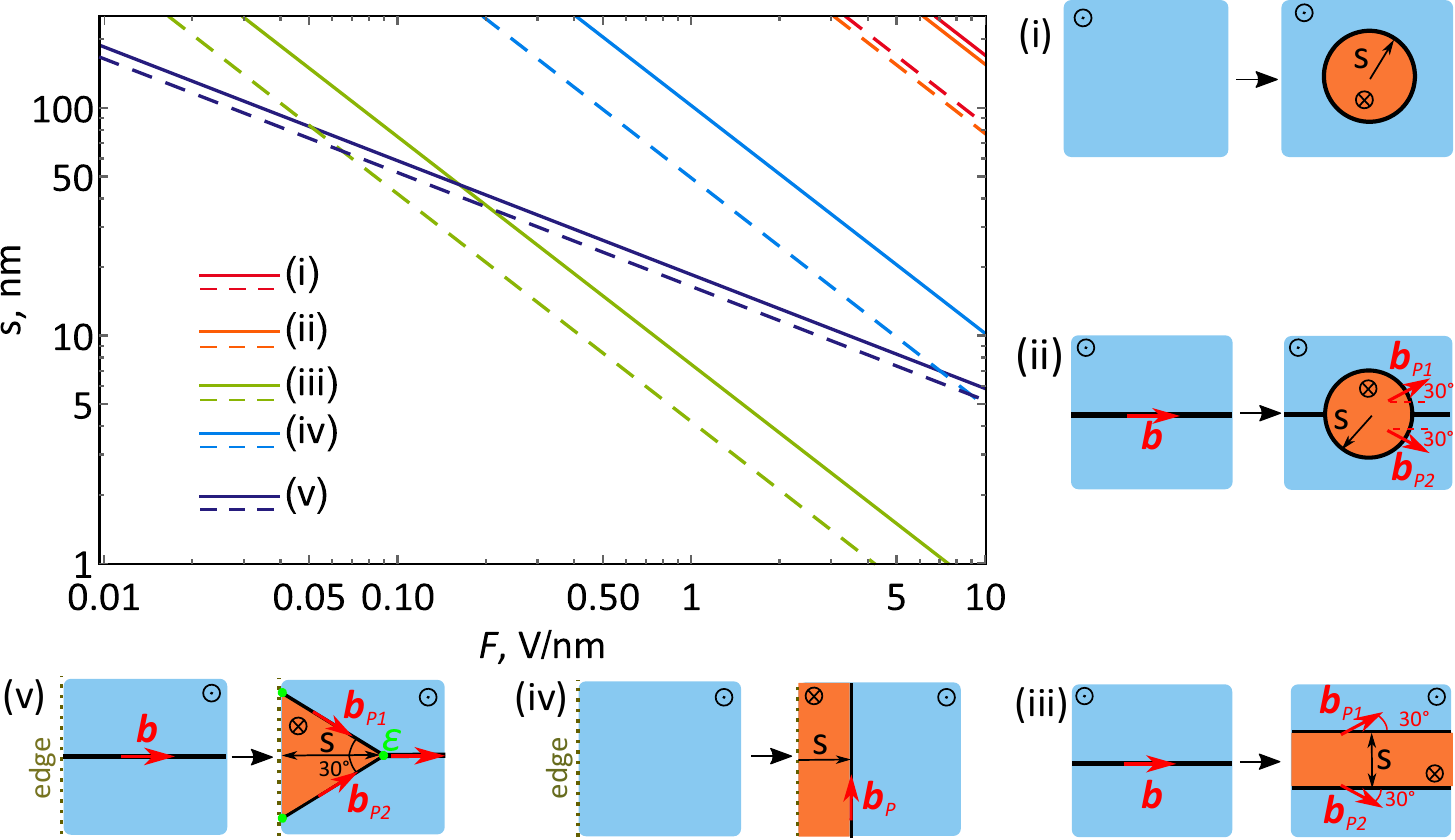}
	\caption{\label{fig:Pswitch} Solid and dashed lines demonstrate critical seed size, Eqs. \eqref{case1}-\eqref{case5}, as a function of out-of-plane electric field for nucleation scenarios (i)-(v) sketched on insets with parameters for P WSe$_2$ and hBN bilayers, respectively.}
\end{figure}
%%%%%%%%%%%%%%%%%%%%%%%%%%%%%%%%%%%%%%%%%%%%

The obtained results for partial dislocations have direct implications for ferroelectric polarisation switch in hBN \cite{woods2021,stern2021} and TMD \cite{Weston2022,wang2021,Ko2023} P bilayers, which, in particular, is associated with nucleation of an oppositely-polarised seed separated by partial dislocation(s) from the main body of domain, followed by the seed area enlargement due to the partial dislocation motion. Having established orientation dependence of partial dislocation energies per unit length for the P bilayers we determine a critical size of a seed, $s$, as a function of out-of-plane electric field, $F$, enabling the seed formation based on energetical reasoning. We consider several scenarios of a seed nucleation sketched in insets of Fig. \ref{fig:Pswitch}. In the scenario (i), circular seed nucleation becomes energetically favourable for the seed radii  
\begin{equation}\label{case1}
	s=\frac{4\mathcal{E}_{P}(0){\bf \rm E}\left(-1-\frac{\lambda}{\mu}\right)}{FP},
\end{equation}  
where numerator comprises orientation-averaged partial dislocation energy per unit length with ${\rm E}(x)$ being the complete elliptic integral  \cite{Prudnikov_Marichev}, and $P$ is areal density of ferroelectric polarisation. The nucleation radius is lowered in scenario (ii), 
\begin{equation}\label{case2}
	s=\frac{4\mathcal{E}_{P}(0){\bf \rm E}\left(-1-\frac{\lambda}{\mu}\right)-\frac{u}{\pi}}{FP},
\end{equation}
where initiation of a seed takes place at a perfect screw dislocation, characterised by Burgers vector $|\bm{b}|=a$ and energy per unit length $u$. In scenario (iii) we examine splitting of a perfect screw dislocation on two parallel partials with Burgers vectors $\bm{b}_{\rm P1,P2}$ misoriented by $30^{\circ}$ with respect to their dislocation lines to satisfy $\bm{b}=\bm{b}_{\rm P1}+\bm{b}_{\rm P2}$. The critical width of emerging seed strip is 
\begin{equation}\label{case3}
s=\frac{2\mathcal{E}_{P}\left(\frac{\pi}{6}\right)-u}{2FP}.		
\end{equation}

\begin{table}%[!t]
	%	\resizebox{\columnwidth}{!}{
	\caption{Parameters of stacking fault energy, $\varepsilon$, perfect dislocation energy per unit length \cite{Enaldiev2022}, $u$, and ferroelectric polarisation density \cite{stern2021,Weston2022}, $P$, used in Eqs. \eqref{case1}-\eqref{case5} and Fig. \ref{fig:Pswitch} for P hBN and WSe$_2$ bilayers. \label{switch_parameters}}
	\begin{threeparttable}
		\begin{tabular}{l|c|c|c}
			\hline
			\hline
			&   $\varepsilon$, eV  & $u$, eV/nm & $P$, $e/\mu$m  \\ 
			\hline 
			P hBN & 0.48 & 1.53 & 3.65 \\ 
			\hline
			P WSe$_2$ & 0.7 & 1.69 & 6.64 \\ 
			\hline
			\hline
		\end{tabular}
	\end{threeparttable}
	%	}
\end{table}

All the above-mentioned scenarios imply seed nucleation inside domain body, whereas it can also form at the sample edge. In scenario (iv), the minimal width of seed nucleating at the edge of a domain via partial screw dislocation inclusion is expressed as follows:
\begin{equation}\label{case4}
	s=\frac{\mathcal{E}_{\rm P}(0)}{2FP}.
\end{equation}   
In an alternative edge-assisted scenario (v), a seed nucleation happens by unzipping segment of a perfect screw dislocation on two partials. Since the energy of two partial screw dislocations of $2s/\sqrt{3}$-length is approximately equal to that of the dissociated segment of perfect dislocation with the length $s$\footnote{Note, that the segment of perfect dislocation splits on two partial screw dislocations according to $\bm{b}=\bm{b}_{\rm P1}+\bm{b}_{\rm P2}$.}, to determine the critical size of a triangular seed we also take into account energy of stacking faults, $\varepsilon$ (indicated by green dots in inset (v) in Fig. \ref{fig:Pswitch}), forming at the ends and crossing of the dislocations. Magnitude of $\varepsilon$ is defined as a difference of adhesion energy densities between stackings of perfect and partial screw dislocations averaged over circular area with diameter equal to width of a partial screw dislocation, Eq. \eqref{Eq:width}. As a result, for nuclei size one obtains:
\begin{equation}\label{case5}
	s=\sqrt{\frac{3\sqrt{3}\varepsilon}{2FP}}.
\end{equation}
In Fig. \ref{fig:Pswitch} we gathered $s(F)$-dependences for all the scenarios with parameters taken for hBN and WSe$_2$ P bilayers, for which low energies of partial dislocations (see Fig. \ref{fig:P_DW}(b)) is expected to facilitate seed nucleation. It appears that for low electric fields (\mbox{$\lesssim0.05\,{\rm V/nm}$} for hBN and \mbox{$\lesssim0.1\,{\rm V/nm}$} for WSe$_2$) the preferential way to nucleate an oppositely-polarised seed is to unzip a segment of perfect screw dislocations on two screw partials at the edge (scenario (v)). However, for such low electric fields, the seed size should surpass $\approx 50$\,nm to be energetically favourable, which makes unlikely the scenario.  In contrast, for stronger electric fields, $\gtrsim 0.1\,{\rm V/nm}$, the seed nucleation is facilitated by splitting of perfect screw dislocation on two parallel mixed partial dislocations (scenario (iii)) along the entire line, with the seed size becoming comparable with partial dislocation widths before electrical breakdown of the bilayers $\sim 3\,{\rm V/nm}$ \cite{Weintrub2022,Weston2022}.

\begin{figure}%[!h]
	%\centering
	\includegraphics[width=1.0\columnwidth]{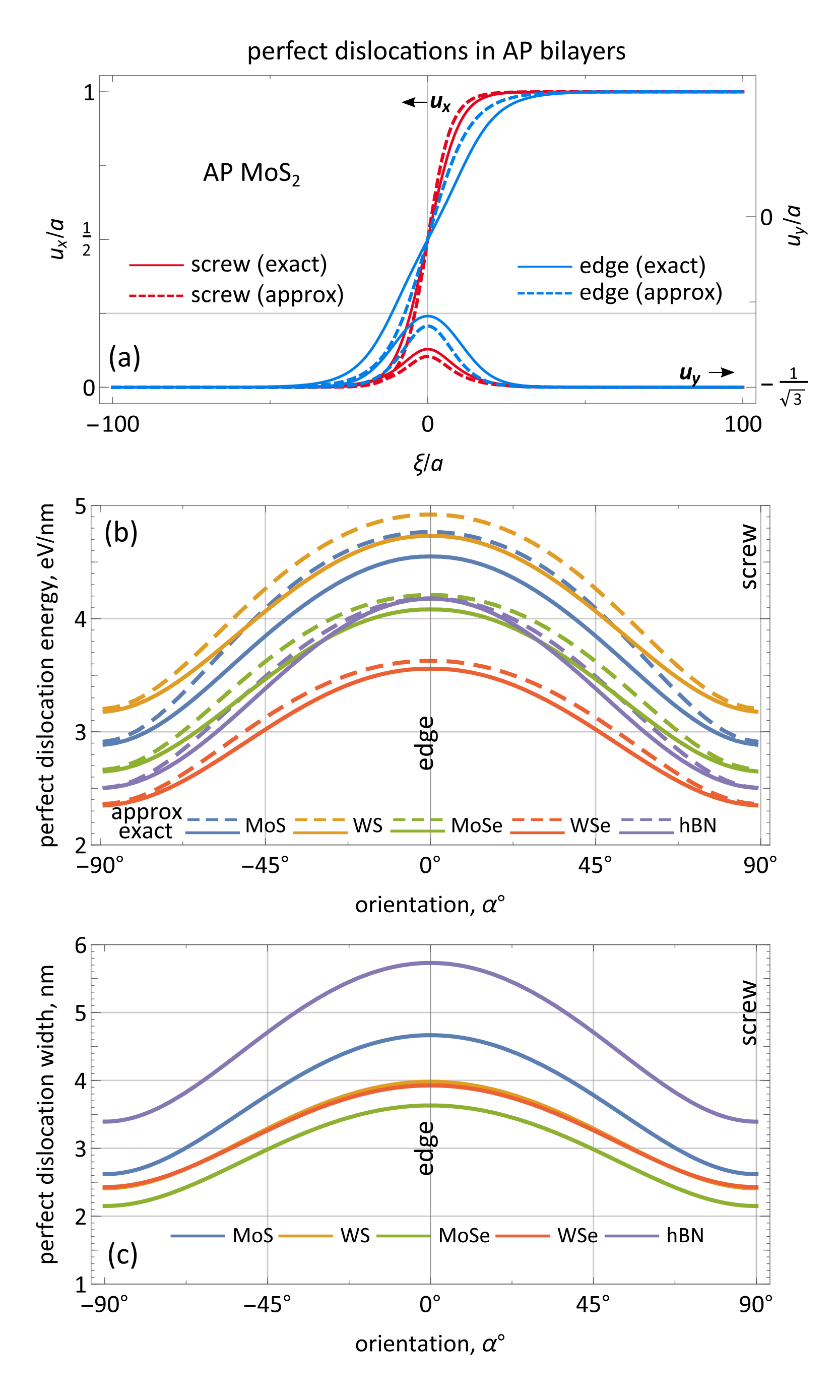}
	\caption{(a) Exact (solid) and approximated (dashed) displacement fields across screw and edge perfect dislocations in AP MoS$_2$ bilayers. (b) Orientation dependences of perfect dislocation energies per unit length for the exact and approximated displacement fields. (c) Orientation dependences of perfect dislocation widths, Eq. \eqref{Eq:widthAP}, for the approximated displacement field. \label{fig:AP_DW}}
\end{figure}

%%%%%%%%%%%%%%%%%%%%%%%%%%%%%%%%%%%%%%%%%%%%%%%%%%%%%%%%%%%%%%%%%%%%%%%%%%%%%%%%%%%%%%%
%%%%%%%%%%%%%%%%%%%%%%%%%%%%%%%%%%%%%%%%%%%%%%%%%%%%%%%%%%%%%%%%%%%%%%%%%%%%%%%%%%%%%%%
{\bf Next, we consider perfect dislocations}, separating identical ABBA stacking domains in AP-bilayers of TMD and hBN. We specify Burgers vector for a single perfect dislocation, $\bm{b}_{AP}=(a,0)$, as shown on the inset in Fig. \ref{fig:0}. Minimizing functional in Eq. \eqref{Eq:1} we obtain the following system: 
\begin{widetext}
\begin{multline}\label{Eq:sys_AP}
	\begin{pmatrix}
		\frac{\mu}{4}+\frac{\lambda+\mu}{4}\cos^2\alpha & \frac{\lambda+\mu}{8}\sin(2\alpha) \\
		\frac{\lambda+\mu}{8}\sin(2\alpha) & \frac{\mu}{4}+\frac{\lambda+\mu}{4}\sin^2\alpha
	\end{pmatrix}
	\begin{pmatrix}
		\tfrac{\partial^2u_x}{\partial\xi^2} \\ \tfrac{\partial^2u_y}{\partial\xi^2}
	\end{pmatrix}=
	%\frac{\partial^2\bm{U}}{\partial\xi^2} =
		\begin{pmatrix}
		G_x\sin\left(G_xu_x\right)
		\left[w_1\cos\left(G_y\widetilde{u}_y-\varphi_1\right)+w_2^{(s)}\cos\left(3G_y\widetilde{u}_y\right)\right] \\
		G_y\cos\left(G_x u_x\right)
		\left[w_1\sin\left(G_y\widetilde{u}_y-\varphi_1\right)+3w_2^{(s)}\cos\left(3G_y\widetilde{u}_y\right)\right]
	\end{pmatrix}
	\\ %\frac{\mathcal{W}_{\rm ad}}{\partial \bm{u}}
	-\begin{pmatrix}
		G_x\sin\left(2G_xu_x\right)\left[w_2^{(s)}+2w_3\cos\left(2G_y\widetilde{u}_y-\varphi_3\right) \right] \\
		2G_y\cos\left(2G_xu_x\right)\left[w_3\sin\left(2G_y\widetilde{u}_y-\varphi_3\right) - w_3^{(a)}\cos\left(2G_y\widetilde{u}_y\right)\right]%\\
		+G_yw_1^{(s)}\sin\left(2G_y\widetilde{u}_y\right)+ G_yw_3\sin\left(4G_y\widetilde{u}_y+\varphi_3\right)
	\end{pmatrix},
\end{multline}
\end{widetext}
where we introduced $(G_x,G_y)=\tfrac{2\pi}{a}(1,\tfrac{1}{\sqrt{3}})$, $\widetilde{u}_y=u_y+\tfrac{a}{2\sqrt{3}}$, \mbox{$w_{1,3}=\sqrt{w_{1,3}^{(s)2}+w_{1,3}^{(a)2}}$}, and $\varphi_{1,3}=\arctan\left(\tfrac{w_{1,3}^{(a)}}{w_{1,3}^{(s)}}\right)$.
The system, Eq. \eqref{Eq:sys_AP} is supplemented with boundary conditions, $\bm{u}(-\infty)=(0,-a/\sqrt{3})$ and $\bm{u}(+\infty)-\bm{u}(-\infty)=\bm{b}_{AP}$, describing crossover between neighboring ABBA domains. Numerical solution of system, Eq. \eqref{Eq:sys_AP}, obtained in Ref. \cite{Enaldiev_PRL} and shown by solid lines in Figs. \ref{fig:0} and \ref{fig:AP_DW}(a), involves variation of both displacement components across screw and edge perfect dislocations. Our aim is to find approximated distributions for the displacement fields, extracting explicit analytical parametric dependences for widths and energies per unit length of perfect dislocations in studied AP bilayers. To this end we notice that the variation range of displacement component orthogonal to $\bm{b}_{AP}$ is confined within interval $a/2\sqrt{3}$, as \mbox{$\max (u_y-u_y(\pm\infty))=u_y(0)+a/\sqrt{3}<a/2\sqrt{3}$}, see Fig. \ref{fig:AP_DW}(a). Therefore, considering \mbox{$(u_y/a)+1/\sqrt{3}$} as a small parameter, we look for an approximated solution setting \mbox{$u_y=-a/\sqrt{3}$} and neglecting $\partial^2u_y/\partial\xi^2$ in the first line of the system, Eq. \eqref{Eq:sys_AP}. This leads to an equation for $u_x$ only, with the following solution: 
\begin{align} \label{Eq:ux}
	u_x(\xi)= 	a-\frac{a}{\pi}\arccot\left(\frac{\sinh\left(\frac{2\xi}{\mathcal{W}_{\rm AP}(\alpha)}\right)}{\sqrt{1-4\frac{w_2^{(s)}-w_3^{(s)}+\sqrt{3}w_3^{(a)}}{w_1^{(s)}-2w_2^{(s)}+\sqrt{3}w_1^{(a)}}}}\right),
%	u_x(\xi)&=\frac{a}{2}\Theta(\xi)+ \\
%	&\frac{a}{\pi}\arctan\left(\left[\frac{\sinh\left(\frac{|\xi|}{\mathcal{W}_{\rm AP}}\right)}{\sqrt{1-4\frac{w_2^{(s)}-w_3^{(s)}+\sqrt{3}w_3^{(a)}}{w_1^{(s)}-2w_2^{(s)}+\sqrt{3}w_1^{(a)}}}}\right]^{\rm sgn(\xi)}\right), \nonumber 
\end{align}
where 
\begin{multline}\label{Eq:widthAP}
\mathcal{W}_{\rm AP}(\alpha) = \\ \frac{a}{\pi\sqrt{2}}\sqrt{\frac{\mu+(\lambda+\mu)\cos^2\alpha}{w_1^{(s)}-6w_2^{(s)}+4w_3^{(s)}+\sqrt{3}\left(w_1^{(a)}-4w_3^{(a)}\right)}}
\end{multline}
is approximated width of the perfect dislocation. To find $u_y(\xi)$ we substitute $u_x(\xi)$, given by Eq. \eqref{Eq:ux} into the second line of system, Eq. \eqref{Eq:sys_AP}, and numerically determine $u_y$ from the resulting equation. The approximated and exact displacement fields across perfect screw and edge dislocations, shown in Fig. \ref{fig:AP_DW}(a) for AP MoS$_2$ bilayers, demonstrate good agreement between each other. Note that the approximation is more accurate for screw-type dislocations as in this case the used condition $|(u_y/a)+1/\sqrt{3}|\ll 1$ is better satisfied. 

In Fig. \ref{fig:AP_DW}(b) we show that energies per unit length of the exact and approximated displacement fields across perfect dislocations differ by less than 5\%  in whole interval of interest $|\alpha|\leq90^{\circ}$ for each AP bilayer. In particular, for screw-type dislocations, where the exact and approximated energies almost coincide, the parametric dependence reads  
\begin{equation}\label{Eq:approxAPenerg}
	\mathcal{E}_{\rm AP}=\tfrac{2a}{\pi}\sqrt{2\mu\left(w_1^{(s)}-2w_2^{(s)}+\sqrt{3}w_1^{(a)}\right)},
\end{equation} 
resulted from Eqs. \eqref{Eq:1}, \eqref{Eq:ux} and $u_y=0$.  

Similar to the partials, widths of perfect dislocations increase from screw to edge orientation, as shown in Fig. \ref{fig:AP_DW}(c), where we plotted the approximated dependences, Eq. \eqref{Eq:widthAP}. Although the latter underestimate the widths of exact displacement fields, they can be used as lower boundary values. %Among studied  AP bilayers, perfect dislocations in hBN are the thickest, as the material possesses the largest elastic and smallest adhesion energy parameters.  

%%%%%%%%%%%%%%%%%%%%%%%%%%%%%%%%%%%%%%%%
%%%%%%%%%%%%%%%%%%%%%%%%%%%%%%%%%%%%%%%%
%%%%%%%%%%%%%%%%%%%%%%%%%%%%%%%%%%%%%%%
%%%%%%%%%%%%%%%%%%%%%%%%%%%%%%%%%%%%%%%

\begin{figure}[!t]
	%\centering
	\includegraphics[width=1.0\columnwidth]{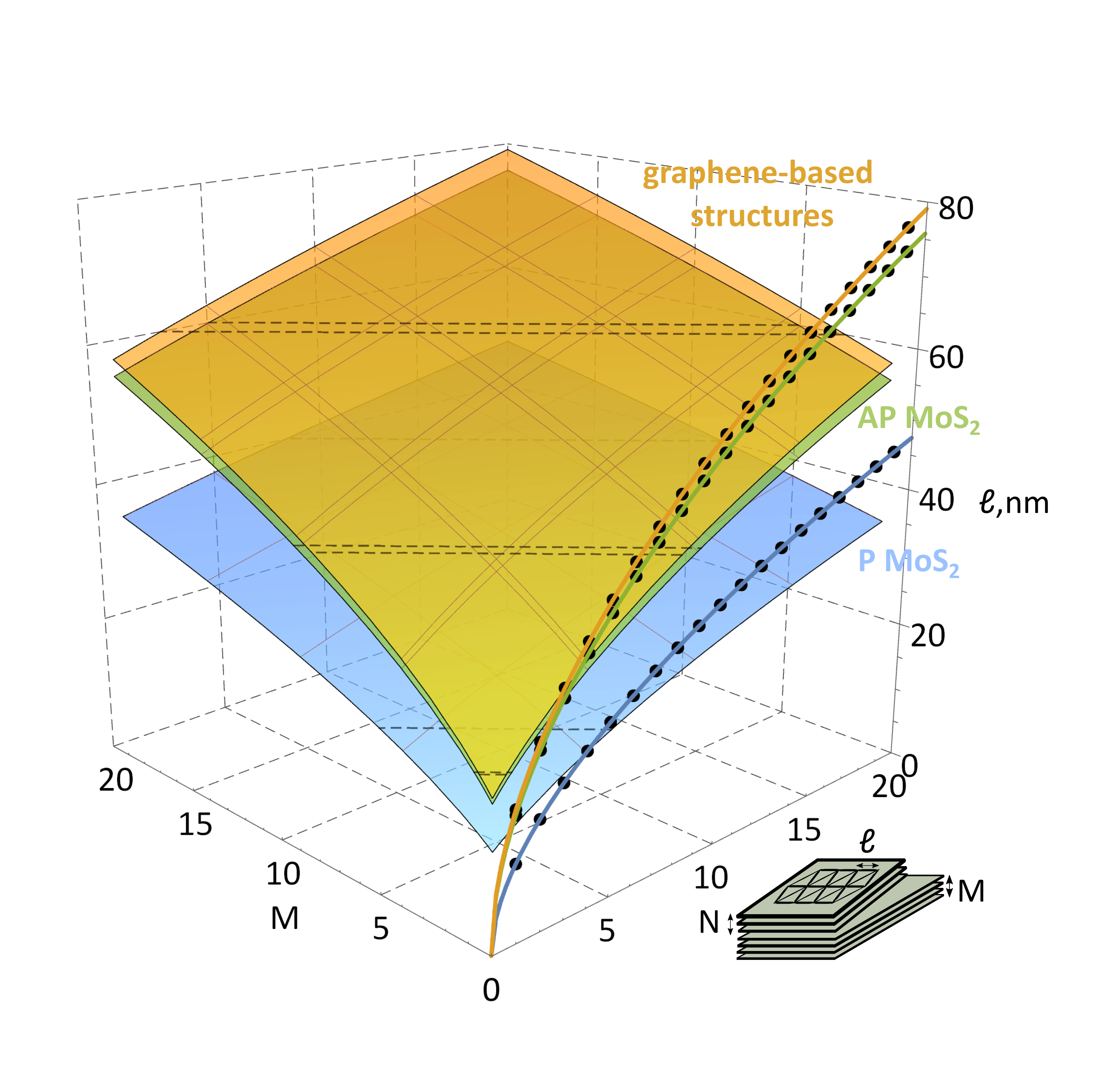}
	\caption{\label{fig:conclusion}  Surface $\ell=\ell_*(N,M)$, Eq. \eqref{Eq:critangle}, divides parameter space $(N,M,\ell)$ characterising each twistronic heterostructures on regions $\ell>\ell_*(N,M)$, featuring relaxation of moir\'e superlattice into domain structure, and  $\ell<\ell_*(N,M)$ with almost non-relaxed moir\'e pattern. For twistronic heterostructures with twisted interface between $N$-layered and bulk crystals parametric space, $(N,\ell)$, is divided along line $\ell=\ell_{*}(2N,0)$.}
\end{figure}

{\bf Finally, we estimate minimal moir\'e superlattice period}, necessary to form dislocation networks at the twisted interface of \mbox{$N+M$}-layered heterostructures sketched on inset in Fig. \ref{fig:conclusion}. Moir\'e superlattices take the form of domain arrays when energy gain from expansion of the preferential stacking areas in moir\'e supercells negotiates energy costs on creation of dislocation networks. The energy costs per supercell, comprising three partial screw dislocations of $\ell$-length (for P) and three perfect screw dislocations of $\ell/\sqrt{3}$-length (for AP) \cite{Enaldiev_PRL}, can be evaluated as $3\,\mathcal{E}_{\rm P}\ell\sqrt{(N+M)/2}$ and $\sqrt{3}\,\mathcal{E}_{\rm AP}\ell\sqrt{(N+M)/2}$, respectively, where $\mathcal{E}_{\rm P/AP}$ is energy per unit length of partial/perfect screw dislocation, Eqs. \eqref{Eq:exactPenergy}, \eqref{Eq:approxAPenerg}, and the factor $\sqrt{(N+M)/2}$ accounts for an increase of the shear module $\mu\to (N+M)\mu/2$ with thickness. The energy gain, $W_{\rm ad}(\bm{u}_0)\sqrt{3}\ell^2/2$, takes into account formation of preferential stacking domains ($\bm{u}_{0}=(0,-a/\sqrt{3})$) with a supercell area. Balance between the costs and gain is attained at the minimal moir\'e period
\begin{equation}\label{Eq:critangle}
	\ell_*^{\rm P/AP} = \frac{\mathcal{E}_{\rm P/AP}\gamma_{\rm P/AP}}{W_{\rm ad}(\bm{u}_{0})}\sqrt{N+M}, \gamma_{\rm P}=\sqrt{6},\gamma_{\rm AP}=\sqrt{2}.
%	\theta^{\rm P/AP}_{*}=\frac{aW_{\rm ad}(\bm{u}_{0})}{2\sqrt{3}\mathcal{E}_{\rm P/AP}\sqrt{N}}.
\end{equation}
In parameter space $(N,M,\ell)$, characterising twistronic heterostructures, Eq. \eqref{Eq:critangle} defines a surface, displayed in Fig. \ref{fig:conclusion} for several heterostructures, which distinguishes a strong relaxation regime of moir\'e superlattice $\ell>\ell_*(N,M)$, associated with formation of domain structures, from a weak relaxation regime, $\ell<\ell_*(N,M)$, where relaxation of moir\'e pattern is negligible. For heterostructures, consisting on $N$-layer stack marginally twisted with respect to bulk crystal, the two regimes are divided by the line $\ell=\ell_*(2N,0)$ (Fig. \ref{fig:conclusion}), where exchange $N\to2N$ assumes no relaxation in the bulk crystal.  

%%%%%%%%%%%%%%%%%%%%%%%%%%%%%%%%%%%%%%%%%%%%%%%%%%%%%%%%%%%%%%%%%
%%%%%%%%%%%%%%%%%%%%%%%%%%%%%%%%%%%%%%%%%%%%%%%%%%%%%%%%%%%%%%
%%%%%%%%%%%%%%%%%%%%%%%%%%%%%%%%%%%%%%%55
%%%%%%%%%%%%%%%%%%%%%%%%%%%%%%%%%%%%%%%%

{\bf Conclusion and Discussion}. We formulated a mesoscale model for dislocations in bilayers of 2D materials combining elasticity theory for crystallitic membranes with an established in the literature \cite{Zhou2015,CarrPRB2018,Enaldiev_PRL} form for interlayer adhesion energy \eqref{Eq:2}. The approach is widely used \cite{CarrPRB2018,Enaldiev_PRL,Koshino2020,Kaliteevski2023} to describe lattice relaxation of moir\'e pattern in various twistronic heterostructures such as small-angle twisted graphene \cite{CarrPRB2018,Koshino2020,Halbertal2023} and MX$_2$ TMD homo- and hetero-bilayers \cite{CarrPRB2018,Enaldiev_PRL,Kaliteevski2023}. We found an exact analytical solution of the model for partial dislocations in P bilayers and determined the dislocation width and energy dependences on orientation and model parameters. Based on this, we considered several scenarios of domain repolarisation in P hBN and TMD bilayers by electric field, establishing preferential way of inversely-polarised seed nucleation by means of dissociation of perfect screw dislocation on two partials. We also note that for bilayer graphene, partial dislocation are equivalent to AB/BA domain walls that sustain quantum confined states \cite{Ju2015,Yin2016,Barrier2024}, for which Eq. \eqref{Eq:4} can be used for more accurate modelling of the state dispersion. Moreover, the orientation dependence of partial dislocation width in bilayer graphene (see Fig. \ref{fig:P_DW}(c)) is in a good agreement with experimentally extracted values \cite{AldenPNAS}. 

For AP bilayers of hBN and TMD we developed a semi-analytical approximation for distribution of displacement field across perfect dislocations, which, for example, results in $\approx3.7$\,nm width for edge dislocations in MoSe$_2$, matching reasonably with experimental value $\approx4$\,nm  \cite{Edelberg2020}. 

In addition, using energy criterium we found parametric conditions for relaxation of moir\'e superlattice into domain structure for multilayer twistronic heterostructures based on graphene and TMD layers, which have recently become in a focus of current research \cite{Halbertal2023,Mullan2023}.

{\bf Acknowledgements.} The author acknowledges fruitful discussions with Roman Gorbachev and Sarah Haigh. The research is supported by the Ministry of Science and Higher Education of the Russian Federation (Goszadaniye) project No FSMG-2023-0011.

\bibliography{refer}

\end{document}